\documentclass{article}

\usepackage{arxiv}

\usepackage[utf8]{inputenc} 
\usepackage[T1]{fontenc}    
\usepackage{hyperref}       
\usepackage{url}            
\usepackage{booktabs}       
\usepackage{amsfonts}       
\usepackage{nicefrac}       
\usepackage{microtype}      
\usepackage{lipsum}

\usepackage{amsmath,amssymb}
\usepackage{subfigure} 
\usepackage[]{algorithm2e}
\usepackage{float}
\usepackage{graphicx,xcolor}
\usepackage{tikz,pgfplots}
\usetikzlibrary{shapes,arrows,shadows,positioning}

\title{Hand bone conduction sound study by using the \\DSP Logger MX 300}

\author{
 Melanie~Adler*, Mariana~Fial\'a S\'anchez*, Constanza~Martini*, Luciana~Vartabedian*, Mat\'ias~Zazzali* \\
  Deparment of Bioengineering, Instituto Tecnol\'ogico de Buenos Aires, Argentina\\
  *All authors contributed equally to this work. The authors are in order of the last name
 \And
  Antonio Quintero-Rinc\'on\\
  Fundaci\'on Lucha contra las Enfermedades Neurol\'ogicas Infantiles (FLENI)\\
  Buenos Aires, Argentina \\
  \texttt{tonioquintero@ieee.org} \\ }

\begin{document}
\maketitle

\begin{abstract}
Bone conduction is the transmission of acoustic energy to the inner ear by different paths involving the bones of the skull. In this work, we use the path the hand provides in order to transmit the sound coming from the cell phone using Bluetooth system. The  aim  of  this  work  was  to  study  the  vibrations  produced  by  a  sound transmitted  through  bone  conduction  between  a  mobile  phone  and the hand  analyzed  with  the  DSP  Logger  MX equipment.\\
\end{abstract}


\section{Introduction}
\label{sec:intro}
Bone conduction is the transmission of subtle sound vibrations to the internal ear through the cranial bones. The devices that are governed by this technology, use this natural process by putting bone conduction transducers in contact with the desired bone \cite{Gelfand2018}. Low frequencies have more acoustical power and generate more vibration than high frequencies \cite{BSA2018}. Therefore, lower frequencies will have a greater effect on the body \cite{integratedlistening}.

Lately, there have been several types of research and product developments trying to implement bone conduction as an alternative to hearing aids \cite{Stenfelt2011}. As the sound vibration travels through the head's tissues, mainly through the bones, sound can get to the cochlea bypassing the external and middle ear. As a starting point, many companies have developed earbuds and headphones implementing this technology, which provides an alternative for those whose traditional hearing aids fail to fully solve their hearing loss. See \cite{Mudry2011,Reinfeldt2015,Eggermont2017} for a complete state-of-the-art of this topic. 

Nowadays, there are different types of hearing aids in the market that use air and bone conduction. Sometimes, in the medical context, depending on the pathology the patient presents, doctors may recommend either one or another. One of the drawbacks that most of these devices have is that they are visible and patients dislike that, even though their hearing is improved. Furthermore, many patients feel that the typical headphones are uncomfortable. Because of this, new models are being created in an attempt to satisfy the patient's needs.

This work is based on a device that instead of being placed on the head or ear, it is placed on the wrist. It consists of a bracelet which receives information via Bluetooth from a mobile phone and converts sound into vibrations using a bone conduction transducer. These vibrations travel from the wrist, through the palm, to the tip of the index finger. Then, by placing the finger near the ear, the person would be able to hear the sound. This device can be used by either a patient who suffers from a pathology in the outer or middle ear or by people who want to use it as a gadget.

A vibration is defined as the oscillating movement a particle does about a fixed point. This movement can either have a regular direction, frequency, and intensity or be, as it usually is, completely random. Therefore, any physical structure, even the human body, can amplify the intensity of vibrations it receives. It is important to know that each body part has a resonant frequency, meaning that vibrations with that frequency are maximized in that region and may induce harmful effects \cite{Soto}.

Vibrations can be measured in many different ways. In this work, we used the SEMAPI DSP Logger MX 300, a device that allows the study of vibrations by measuring changes in acceleration, speed or movement of a surface. The team chose to focus the study on acceleration, which cause the software to give more weight to higher frequency components \cite{measuringvibration}.
SEMAPI DSP Logger MX 300 can measure any signal between 4-20 mA from 0-10 V or 0-5 V at any temperature. Regarding vibrations, it incorporates spectrums of 400, 2000 and 4000 lines of resolution. This device is usually used for the collection of condition status monitoring data, the analysis and correction of the root cause of a problem and rotating equipment applications, such as motors, pumps, fans, gearboxes and other rotating machinery composed of bearings. Nevertheless, in this work, it was used to measure the vibrations in some specific points of the palm of the hand, evaluated in ten different subjects, which helped to characterize the vibration conduction. 

The rest of this paper  is  organized  as  follows: Section \ref{sec:dsp} introduced the DSP Logger MX 300. Section \ref{sec:meth} describes the methodology used. Section \ref{sec:res} explains the results obtained which are discussed. Finally, in Section \ref{sec:con} conclusions and future work are reported.

\section{DSP Logger MX 300}
\label{sec:dsp}
 This device can be used to measure multiple variables. It has a digital signal processor, analog/digital converters of 16 and 24 bits and uses a 16 MB FLASH memory which manages and splits according to each program, see Table \ref{table 1}.

\begin{table}[htbp]
\caption{Flash memory division for each DSP Logger MX 300 program \cite{UserManual}}
\label{table 1}
\centering
\begin{tabular}{|l|l|}
\hline \hline
Quantity & Data File \\
\hline \hline
12MB & Data Collector \\ 
\hline
512 KB & Phase Analysis \\ \hline
512 KB & Machine Balancing \\ \hline
1MB & Measure out of routes \\
\hline
512 KB & Auxiliary Measurements \\
\hline \hline
\end{tabular}
\end{table}

The DSP Logger MX 300 has 5 types of programs: Data Collector, Phase Analysis, Machine Balancing, Measure Out of Route and Auxiliary Measurements. For this work, it was used only in the Measure Out of Route mode. This program allows recording any vibration measurement, whether in the form of a wave, global values or spectra. The amount of measurements is limited by the free memory capacity of the equipment. 

It should be noted that a measurement can be made through two channels; one measures the acceleration and the other the speed. In our case, giving that we were only interested in the acceleration, we used just one channel. Apart from that, the Measure Out of Route program allows editing and deletion of measurements already made.

To start a new measurement, the device needs to be configured taking into account certain parameters:

\begin{itemize}
    \item Measurement: The number of lines of definition wanted in the spectrum. In our case it was 4000.
    \item Variable to measure: Variable to measure within the possible ones (acceleration, velocity, envelope or displacement). Acceleration was chosen in this case.
    \item Maximum frequency: This parameter is to set the cutoff frequency of the measurement. It was 500 Hz for this work.
    \item Channel: Selection of the input channel
    \item Gain: Allows to toggle between X1 gain or X10 gain.  In our case it was X10 to get a better resolution.
\end{itemize}

There are more parameters such as windows, filters, threshold, slope, among others, which were not taken into account.

Once all the parameters had been configured, measurements were made. Finally, after performing all the measurements, the data obtained was sent to a computer to plot the graphics obtained with the DSP Logger MX 300 (Section III). 
 
\section{Methodology}
\label{sec:meth}

\begin{figure}[!h]
	\centering
	\subfigure[]{\includegraphics[clip,width=0.45\columnwidth]{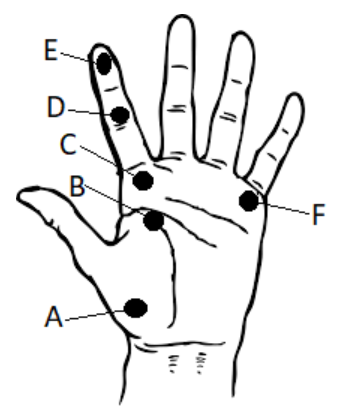}\label{FIG:fig1a}}
	\subfigure[]{\includegraphics[clip,width=0.45\columnwidth]{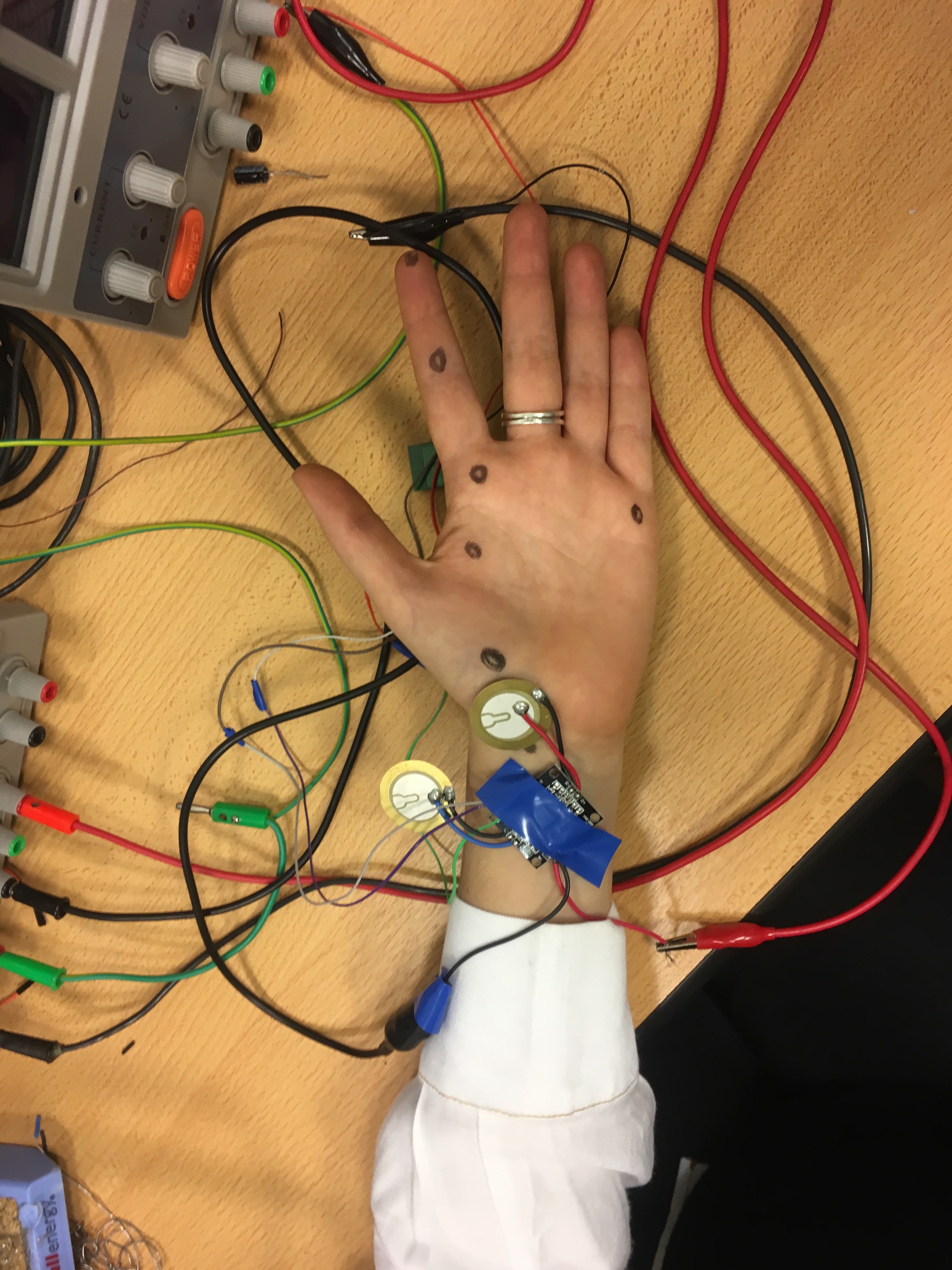}\label{FIG:fig1b}}
	\caption{(a) Spots were the sensor of vibration was positioned. (b) Piezoelectric setting during the experiment.}
	\label{fig:hand}
\end{figure}

With the aim of studying the vibrations produced in the hand due to bone conduction, a prototype of a hearing device was placed on the subject's wrist and measurements were carried out at six different parts of the hand (as shown in Figure \ref{fig:hand}). The group of the study consisted of ten different people, five men, and five women, all in the 22-23 age range.

\begin{figure}[!h]
		\centering
		\tikzstyle{none} = [above, text centered]
		\tikzstyle{block} = [draw, fill=white, text centered, minimum height=2.5em, minimum width=2.5em,drop shadow]
		\tikzstyle{block1} = [block, fill=blue!20, minimum width=3em]
		\tikzstyle{block2} = [block, fill=red!20,  minimum width=6em, minimum height=3em, rounded corners]
		\begin{tikzpicture}[
			->,>=stealth',auto,thick,
			node distance=1.5cm,
		]
			\node[block1] (N1) {Cell Phone App};
			\node[block2] (N2) [below of=N1] {Sound};
			\node[block1] (N3) [below of=N2] {Bluetooth};
			\node[block2] (N4) [below of=N3,text width=11em] {Bone Conduction Device};
			\node[block1] (N5) [right=4cm of N1.north] {Transducer};
			\node[block2] (N6) [below of=N5, text width=7em] {Hand};
			\node[block1] (N7) [below of=N6] {Transmission};
			\node[block2] (N8) [below of=N7,text width=7em] {Accelerometer};
			\node[block1] (N9) [below of=N8] {DSP Logger};
			
			\path[every node/.style={font=\sffamily\small}]
				(N1) edge node [right] {} (N2)
				(N2) edge node [right] {} (N3)
				(N3) edge node [right] {} (N4)
				(N4) edge[out=0, in=180] node [left] {} (N5)
				(N5) edge node [left] {} (N6)
				(N6) edge node [left] {} (N7)
				(N7) edge node [left] {} (N8)
				(N8) edge node [left] {} (N9)
				;
		\end{tikzpicture}
\caption{Circuit diagram  used in the methodology proposed.}
\label{fig:circuit}
\end{figure}
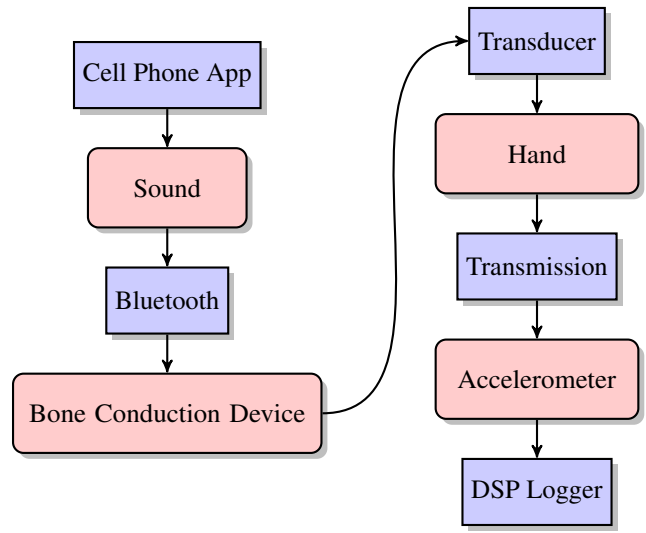

The bone conduction device included a piezoelectric transducer, a PAM8610 10W Stereo Class-D Audio Power Amplifier \cite{datasheetPAM} and a Bluetooth module XS3868 \cite{core2018}. This device was connected -via Bluetooth- to a cell phone, from which an audio signal was emitted. The app used to generate the signal was Frequency Sound Generator by LuxDeLux available both for iOS and Android \cite{FSG} and the signal generated was a pure tone at 440Hz. A diagram of the whole circuit is shown in Figure \ref{fig:circuit}. 'A440' is the name given colloquially to the sound that produces a 440 Hz vibration and serves as a general tuning standard for musical pitch perception \cite{Gelfand2018}.
For this purpose, a pure tone with a frequency of 440 Hz was used in order to determine if the transmission was being carried out correctly in the different points of the hand that were selected for this study, see Figure \ref{fig:hand}.

Vibrations were measured at points A, B, C, D, E and F (shown in Figure \ref{FIG:fig1a}) using the DSP Logger MX 300. In order to get a reference signal, vibration coming out directly from the transducer was also measured. All the gathered data was then transferred to a PC and analyzed using the SEMAPI DSP Data Management software. Posterior processing was performed in MATLAB in order to obtain better and clearer statistical results shown in Figures \ref{fig:lossresult} and \ref{fig:transmissionresult}. 

\section{Results and discussion}
\label{sec:res}
In this section, the proposed methodology was evaluated by studying the vibrations, measuring the changes in the acceleration ($g$) using the DSP Logger MX 300 equipment introduced in Section \ref{sec:dsp}.
A healthy and young ear is sensitive to frequencies between 20 Hz and 20 kHz. However, this margin varies according to each person and decreases with age \cite{Gelfand2018}. 

Figure \ref{fig:test} shows, as an example, the result obtained with the DSP Logger MX 300 for one of the subjects. It can be seen that most of the signal power is concentrated at 440 Hz, as it was expected. However, there are peaks at 50 Hz and its multiples, which were initially thought to be caused by power line noise and its harmonics. In order to verify this assumption, the spectrum of noise was measured by hanging the sensor in the air. The result of this measurement can be seen in Figure \ref{fig:noise}, where peaks at 50Hz and its harmonics are visible and are as strong as in measurements made in subjects and the reference signal, which can be seen in Figure \ref{fig:ref}.  In order to evacuate doubts, further studies will be done isolating the experiment from any source of power line noise. 

In terms of functionality, points A and E can be said to be the most relevant due to their proximity to the transducer-hand and hand-head interface regions, respectively. Knowing more about these points may help decide whether a bracelet-like or a ring-like device would be more favorable in order to minimize signal power loss.

\begin{figure}[!h]
\centering
\includegraphics[clip,width=0.63\columnwidth]{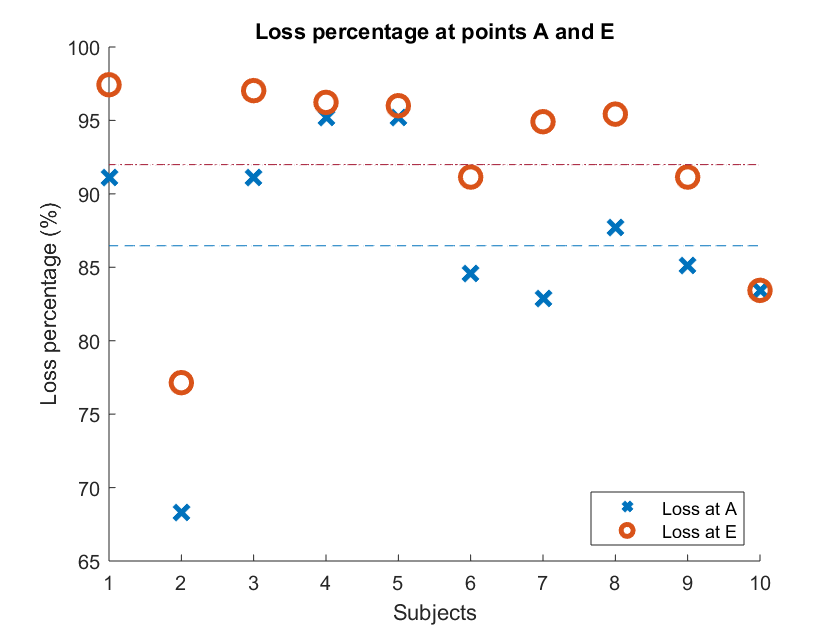}
\caption{Signal power loss ($1-\frac{P_{out}}{P_{ref}}$) between the reference point (piezoelectric vibration) and points A and E for subjects 1 to 10. Dashed lines show the average loss at point A (blue) and E (red).}
\label{fig:lossresult}
\end{figure}

Figure \ref{fig:lossresult} shows how much of the initial signal power is lost at points A and E. The average loss for all subjects is 86\% and 92\% for points A and E, respectively. Although there is a difference between them, it can be inferred that most of the power loss takes place at the transducer-hand interface.
It should also be noted that the attenuation of the vibration depends on each person and its tissues, some people have larger conservation of signal power that other, although this was not the aim of this study.

\begin{figure}[!h]
\centering
\includegraphics[clip,width=0.63\columnwidth]{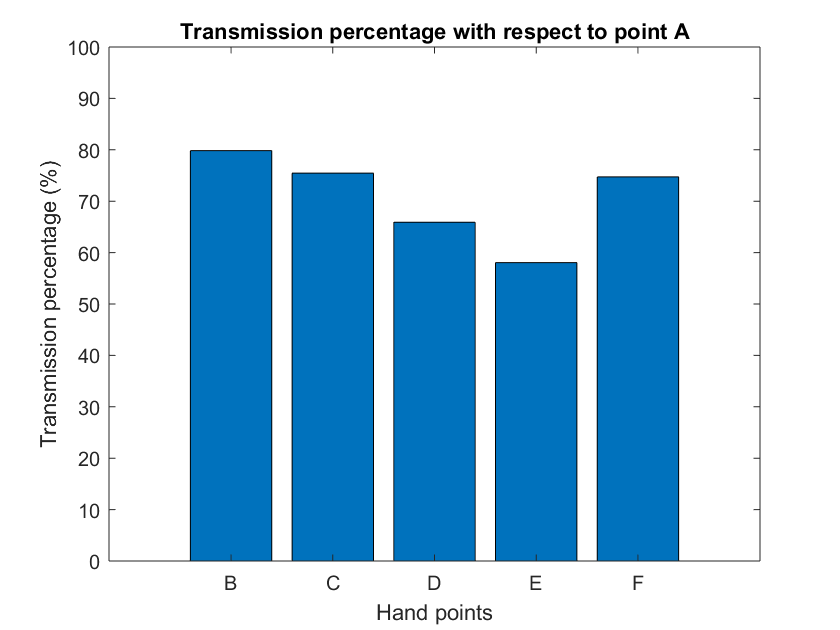}
\caption{Signal power at points B, C, D, E and F relative to signal power at point A ($\frac{P_{out}}{P_A}$).}
\label{fig:transmissionresult}
\end{figure}

\begin{figure}[!h]
\centering
\includegraphics[clip,width=0.63\columnwidth]{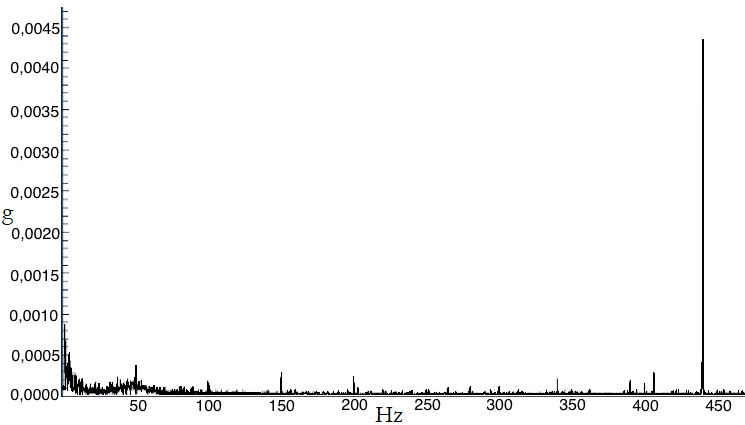}
\caption{Result of one of the subjects showing that the signal power is concentrated at 440 Hz.X-axis shows frequency up to 500Hz; Y-axis shows vibration measured as acceleration in Gs.}
\label{fig:test}
\end{figure}

\begin{figure}[!h]
\centering
\includegraphics[clip,width=0.63\columnwidth]{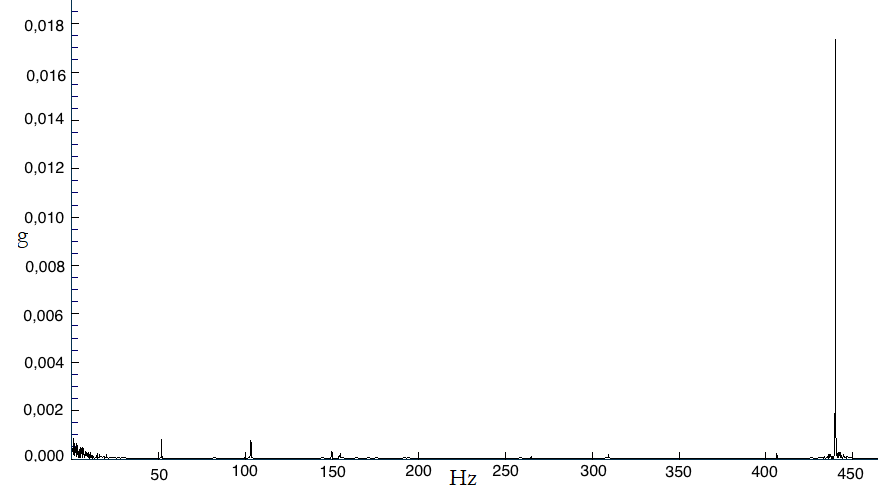}
\caption{Spectrum of the reference signal. X-axis shows frequency up to 500Hz; Y-axis shows vibration measured as acceleration in Gs.}
\label{fig:ref}
\end{figure}

\begin{figure}[!h]
\centering
\includegraphics[clip,width=0.63\columnwidth]{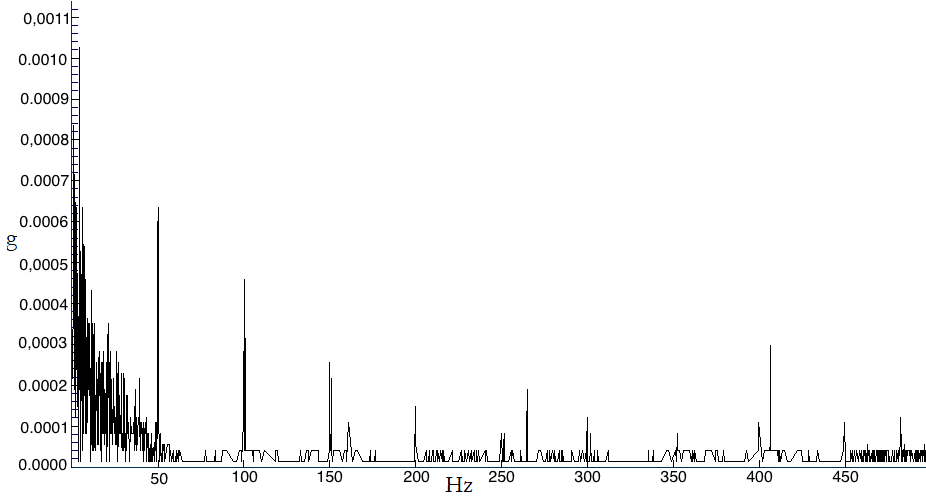}
\caption{Base of noise that appears in every measurement. Demonstration that peaks at 50 Hz and its harmonics are due to power line noise since it was measured without contact with any surface. X-axis shows frequency up to 500Hz; Y-axis shows vibration measured as acceleration in Gs.}
\label{fig:noise}
\end{figure}

Figure \ref{fig:transmissionresult}, on the other hand, shows how much of the signal power that gets to point A is transmitted to the rest of the points. There is a clear pattern showing that signal power decreases with distance. Power at point F being similar to power at point C empowers this assumption since the distance to point A is comparable between them.
This experimentation study allowed to determine the best possible position in order to use a bone conduction device in the hand.

\section{Conclusions}
\label{sec:con}

A study that allows knowing both the quality of transmission as well as the attenuation caused by the hand facilitates the design of a future communication device using bone conduction. With the acquired results we can conclude that, in terms of sound transmission, a ring design would have less power loss than a bracelet design since it decreases with distance. However, the transducer, battery and other electronic components to be used -such as an amplifier or analog filter- would have to be much smaller, therefore causing implementation difficulties \cite{Brooke2009}. In addition, the inclusion of a microphone may bring extra problems as it would be distant from the mouth.

In order to continue the research, hand bone conduction sound study will proceed with the investigation of different transducers \cite{Frohlich2018}. Aiming to achieve the best sound transmission through the hand, the selection of these transducers will be on the basis that they are specifically made for bone conduction. 

Further research should be done analyzing a greater range of frequencies \cite{Popelka2009}, body resonant frequencies \cite{Hakansson1994}, and device calibration \cite{Margolis2014}. Also, in order to obtain results with more feedback from subjects, performing audiometric tests comparing traditional bone conduction devices with bracelet and ringbone conduction devices would be useful \cite{Frohlich2018}.

In spite of the issues explained before, it has been shown that these novel bone conduction designs, either bracelet or ring-like devices, have plenty of potential and could benefit a lot of people. 

\section*{Acknowledgement}
We are grateful to Eng. Nicolas Oyarzabal from Centro de ingenier\'ia de los Materiales (CeMat) laboratory from ITBA, Argentina, by the SEMAPI MX 300 vibration measurement equipment.\\
This study was conducted when AQR  worked at the Buenos Aires Institute of Technology (ITBA) during the  "Initiation to research and technological development" (I+D2018).

\bibliographystyle{IEEEtran}

\end{document}